# Influence of Diaphragm Dynamics on Shock Wave Propagation in Double-Diaphragm Shock Tubes

Touqeer Anwar Kashif*, Janardhanraj Subburaj[†], Md Zafar Ali Khan [‡] and Aamir Farooq [§]
*Mechanical Engineering, PSE Division, King Abdullah University of Science and Technology (KAUST), Thuwal, 23955-6900, Kingdom of Saudi Arabia*

**Shock tubes have become indispensable tools for advancing research across diverse fields such as aerodynamics, astrophysics, chemical kinetics, and industrial applications. The present study examines shock velocity variations in double-diaphragm shock tubes and their impact on flow dynamics at the end of the shock tube. Experiments were conducted using helium as the driver gas at 30 bar and argon as the driven gas at 100 Torr, with two different mid-section pressures (4.8 and 18.8 bar). A previous study highlighted two mechanisms governing the shock velocity profile: a higher peak shock velocity in the case of lower mid-section pressure due to stronger shock heating of the mid-section gas, and a second stage of acceleration of the shock wave resulting from the higher mid-section pressure and delayed opening of the first diaphragm. Pressure histories of the mid-section between the two diaphragms and high-speed imaging of the diaphragm-opening process helped validate the mechanisms influencing shock velocity. Additionally, numerical simulations were performed (i) to understand the effect of diaphragm-opening dynamics on the experimental shock profiles and (ii) to quantify the axial temperature as well as pressure profiles within the shocked gas. The simulations revealed that while axial pressure variations were not significantly affected by shock velocity variation, there were large temperature gradients in the shocked gas, with up to 6.5% higher peak temperature in the case of lower mid-section pressure compared to its counterpart. These pivotal findings underscore the need to control the mid-section pressure to minimize temperature variations within the shocked gas region.**

## Nomenclature

*PD*      =    primary diaphragm, in *bar*

*SD*      =    secondary diaphragm in, *Torr*

---

*PhD student, Mechanical Engineering, touqeer.kashif@kaust.edu.sa, AIAA Student Member
[†]Research scientist, Mechanical Engineering, janardhanraj.subburaj@kaust.edu.sa, AIAA Professional Member, Corresponding author
[‡]PhD student, Mechanical Engineering, muhammed.khan.1@kaust.edu.sa
[§]Professor, Mechanical Engineering, aamir.farooq@kaust.edu.sa

| | | |
|---|---|---|
| $P_4$ | = | pressure of driver gas |
| $P_1$ | = | pressure of driven gas |
| $\gamma_4$ | = | specific heat ratio of driver gas |
| $\gamma_1$ | = | specific heat ratio of driven gas |
| $a_1$ | = | speed of sound of the driven gas, in *m/s* |
| $a_4$ | = | speed of sound of the driver gas, in *m/s* |
| $D$ | = | diameter of the driven section of the shock tube, in *m* |
| $X/D$ | = | driven length normalized by diameter |
| $h$ | = | orifice opening at the diaphragm location, in *mm* |
| $M_s$ | = | shock Mach number |
| $ISW_{mid}$ | = | incident shock wave in the mid-section |
| $ISW_{driven}$ | = | incident shock wave in the driven section |
| $T_2$ | = | temperature of shocked driven gas, in *K* |
| $P_2$ | = | pressure of shocked driven gas, *in bar* |
| $T_{2,endwall}$ | = | temperature of shocked driven gas near the driven endwall, in *K* |
| $P_{2,endwall}$ | = | pressure of shocked driven gas near the driven endwall, in *bar* |

## I. Introduction

Shock tubes play an important role in research in various disciplines such as aerodynamics, astrophysics [1], chemical kinetics [2, 3], material science [4, 5], biological applications [6], and numerous industrial applications [7]. A comprehensive review by Balan et al. [5] and Subburaj et al. [8] explores various design adaptations of shock tubes for diverse applications. In its simplest form, a shock tube typically consists of a driver and a driven section, separated by a diaphragm [1]. Typically, the driver section is filled with a light gas, such as hydrogen or helium, while the choice of driven gas varies depending on the specific needs of the research, ranging from atmospheric compositions for aerodynamic studies to fuel mixtures for combustion research. The materials used as diaphragms differ according to the required pressure levels, incorporating lighter materials like Mylar or polycarbonate for lower pressures, and metals such as aluminum, copper, or stainless steel for high pressure scenarios [3, 9]. The rupture of the diaphragm generates a shock wave as the driver gas expands into the driven section. In ideal shock theory, it is assumed that the diaphragm ruptures instantaneously and a shock wave is formed at the location of the diaphragm. In addition, a one-dimensional, inviscid and adiabatic flow is assumed inside the shock tube. Furthermore, the shock strength can be calculated from the initial conditions in the driver and driven sections, using these assumptions. The equation relating these parameters are shown in Eq. 1.

$$P_{41} = \frac{2\gamma_1 M_S^2 - (\gamma_1 - 1)}{\gamma_1 + 1} \left(1 - \frac{\gamma_4 - 1}{\gamma_1 + 1} \frac{a_1}{a_4} \left(M_S - \frac{1}{M_S}\right)\right)^{\frac{-2\gamma_4}{\gamma_4 - 1}} \quad (1)$$

where the subscript 4 indicates driver region and subscript 1 indicates driven region. $P_{41}$ stands for driver-to-driven pressure ratio, $a$ is the local speed of sound, $\gamma$ is the specific heat ratio, and $M_S$ is the shock Mach number.

However, the rupture of the diaphragm is not instantaneous in practical scenarios and occurs over a finite time, which slows the expansion of the driver gas in the driven section and leads to a weaker initial shock wave as compared to that predicted by Eq. 1). Subburaj et al. [10] showed that the initial stages of diaphragm opening, including the rate and profile of opening, play crucial roles in resulting incident shock Mach number and test time. White [9] demonstrated that the shock front accelerates until the diaphragm fully opens to match the internal cross-sectional area of the shock tube, at which point it reaches peak velocity. The shock formation distance ($x_f$) describes the extent of this acceleration phase. Therefore, in reality, the driven section must be designed considering the shock formation distance under different operating conditions. Several other studies have attempted to quantify the shock formation distance by correlating it with the opening time of the diaphragm [11–14]. Moreover, beyond the initial acceleration phase, the shock wave begins to attenuate with the growing boundary layer behind the incident shock front. Mirels' theoretical contributions [15–18] offer a framework for calculating the thickness of the boundary layer under laminar and turbulent flow assumptions, as well as their effect on test time duration and shock strength. Moreover, several theories predicting the boundary layer thickness and its influence on shock attenuation have been extensively discussed, with a comprehensive report provided by Spence [19].

With respect to computational efforts in this area, shock tube simulations have historically underrepresented the combined effects of diaphragm opening and boundary layer interactions. An early study by Petrie-Repar et al. [20] was among the first to model the diaphragm-opening process as a dilating iris, simulating the experiments of Miller and Jones [21] with a fixed 200 $\mu$s opening time and an axisymmetric approximation that resulted in under-predicted velocities due to a reduced flow area. Gaetani et al.[22] used inviscid flow simulations to develop a correlation predicting the reduction in shock strength with a decrease in flow area at the diaphragm orifice. Lamnaouer et al.[23] performed viscous simulations with turbulent boundary layer modeling without incorporation of diaphragm opening. Satchell et al. [24] introduced a laminar viscous shock tube solver with enhanced shock tracking and modeled the diaphragm as a dilating iris to validate experimental data for various Mach numbers. This approach provided insights into the thermodynamic changes behind moving shocks[25]. Andreotti et al. [26] performed extensive shock-tunnel simulations that included fluid-structure interactions and achieved a close match between 3-D and 1-D simulations by modeling gradual diaphragm openings. However, they did not explore the shock wave formation or attenuation process in detail.

The aforementioned studies focus predominantly on single diaphragm shock tubes. Double diaphragm shock tubes, which include a mid-section between the driver and driven sections, offer enhanced control and flexibility in operation [27]. In their design, the driver and mid-section are separated by a primary diaphragm (PD), while the mid-section and driven section are separated by a secondary diaphragm (SD). A concise overview of the operation of a double-diaphragm shock tube is provided here, given its relevance to the current paper. Initially, the driven section is filled with the test gas to a specified pressure. Subsequently, the driver and mid-sections are pressurized simultaneously by controlling a pneumatic valve. Once the pressure in the mid-section reaches a predefined ratio relative to the driver pressure, the valve is closed to isolate the driver section for further pressurization. The operation continues with the venting of the mid-section to facilitate the rupture of the PD, followed by the rupture of the SD, ultimately generating a shock wave in the driven section. Additional details regarding the operation of double-diaphragm mode has been provided by Kashif et al. [28].

Although double-diaphragm shock tubes offer enhanced control and flexibility, the understanding of shock dynamics and non-uniformities in the shocked gas region remains limited. Notable contributions have been made by Miller and Jones [21], who demonstrated shock velocity measurements for various driven gases, achieving speeds exceeding 2000 m/s. More recently, Kashif et al. [28] identified two shock velocity profiles based on the pressure conditions in the mid-section of double-diaphragm shock tubes. They proposed that these variations were influenced by factors such as the jump in the thermodynamic conditions in the mid-section post-rupture of PD and the incomplete opening of PD. However, these mechanisms have not yet been validated experimentally, presenting an opportunity for further investigation into the effects of double diaphragm configurations on shock dynamics. The recent numerical study by Currao et al. [29] represents a significant advancement in the simulation of double-diaphragm shock tubes, incorporating both 2-D and 3-D numerical models. They demonstrated the feasibility of accurately capturing shock velocity trends by modeling the dynamics of diaphragm movement as a dilating iris. While the experimental validation of this study was confined to the properties at the end of the shock tube, there was no discussion of the initial phases of shock propagation. Notably, this work identified a second stage of acceleration, in experiments and numerical simulations, as depicted in Figure 21 of their publication [29]. This acceleration phase, similar to the observations made by Kashif et al.[28], indicates the need for a deeper understanding of underlying mechanisms and broader implications on non-idealities in shock dynamics.

Given the limited investigations on double-diaphragm shock tubes, the present work aims to delve deeper into the phenomena of shock velocity variations due to successive rupture of two diaphragms and their implications on shock dynamics. To address these gaps, the present study is structured to provide both experimental and numerical insights through the following objectives.

- **Diaphragm opening imaging and shock velocity measurements:** This study extends the observations of Kashif et al. [28], incorporating advanced diagnostics to assess the variation of shock velocity and diaphragm opening processes.

- **Numerical simulation of double-diaphragm shock tube:** By replicating the diaphragm opening profiles

observed in experiments, this research aims to align numerical simulations with measured data, enhancing our understanding of the complex flow phenomena within shock tubes.

- **Effects of diaphragm opening on the shocked gas:** The study also investigates how variations in shock velocity influence the axial temperature and pressure profiles within the shocked gas, aiming to elucidate the thermodynamic impacts of these variations.

This paper is organized into several sections to describe the methodology and findings. Section II details the experimental and numerical methodologies used in the study. Section III discusses the experimental results that validate the mechanisms proposed in [30] and present the validation of our numerical results with experiments. Section IV explores complex flow dynamics in the double-diaphragm shock tube operation, including a discussion on axial variations in temperature and pressure. Finally, Section V concludes the paper with a summary of the findings and scope for future research.

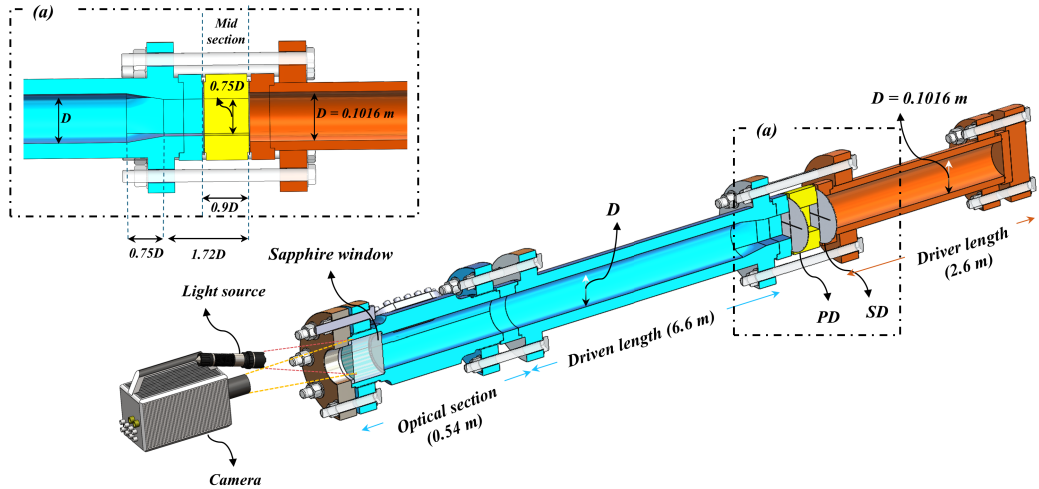

Fig. 1  Schematic of HPST facility showing the dimensions of various sections and visualization setup to image the diaphragm opening process. The images of the SD's opening is recorded by a high-speed camera placed at the driven end wall and the diaphragm is illuminated by a light source placed near the camera.

## II. Methodology

### A. Experimental Methodology

Experiments were conducted in the High-Pressure Shock Tube (HPST) at the King Abdullah University of Science and Technology (KAUST), Saudi Arabia. This facility, made of stainless steel, is engineered to withstand static pressures up to 300 bar. The experimental setup includes a driven section measuring 7.14 meters and a driver section of 2.6 meters, both with an internal diameter of 101.6 mm. The last 0.54 meters of the driven section comprises an optical section equipped with a sapphire endwall, that allows visualization of the diaphragm opening process. Figure

| Notation | Driver Pressure ($P_4$), bar | Mid-section's pressure | | Driven Pressure, ($P_1$), Torr | $P_{41}$ | $M_{s,endwall}$ |
|---|---|---|---|---|---|---|
| | | $P_{mid,ini}$ | $P_{mid,rupture}$ | | | |
| C1 | 30 | 8 | 4.8 | 100 | 225 | 4.13 |
| C2 | 30 | 23.5 | 18.8 | 100 | 225 | 4.14 |

**Table 1  Operating conditions in the various sections of the double-diaphragm shock tube.**

1 shows a schematic cross-section of the HPST, highlighting the capability of the tube to function in either single- or double-diaphragm mode; this study, however, exclusively employed the double-diaphragm configuration. The mid-section, as indicated in Fig. 1a, has a square cross-section and the facility features a transition from a square to a round section to ensure smooth flow dynamics. Helium was used as the driver and mid-section gases and argon was used as driven gas, sourced from AirLiquide Inc. at a purity of 99.999 %. The gauge pressures in the driver section were adjusted to 30 bar, and the driven section pressures were set to 100 (0.133 bar), as detailed in Table 1. Aluminum diaphragms, 2.1 mm thick, were scored in an 'X' pattern to guarantee a clean and symmetrical opening.

The trajectory of the shock wave was monitored using twelve pressure transducers (Model: 112B05, PCB Piezotronics, USA) located at distances between 0.19 m and 6.59 m from the diaphragm location. The pressure sensors were shielded with a silicone RTV coating for thermal protection and connected to a data acquisition system (National Instruments, USA) with a response time of 1 $\mu$s. The accuracy of the shock velocity measurements was within 1 %. For further details on HPST's capabilities and uncertainty analysis, readers are referred to additional sources [31–33]. The measured shock velocity serves as the basis for calculating key thermodynamic properties, such as temperature and density, behind both the incident and reflected shock waves, using ideal shock relations. A high-speed camera (Model: SA-X2, Photron Limited, Japan) captured the motion of the diaphragm's petals at 150,000 frames per second, with a temporal resolution of 6.6 $\mu$s and a pixel resolution of about 1 mm in both the *x* and *y* axes. Equipped with a 70-200 mm lens, the camera was positioned about 8 m away from the diaphragm's location for optimal visualization. The acquired images were processed in MATLAB using a Canny edge detection filter to delineate the edges of the diaphragm petals, followed by binarization to estimate the area of aperture opened. During instances of minimal aperture opening, manual tracking was necessary due to limitations in automatic edge detection capabilities. Using the described imaging setup, the opening of the SD could be viewed completely, while only a portion of the PD's opening could be observed. Further details on the diaphragm opening process has been provided in the results section.

## B. Numerical Methodology

The computational domain consists of a 2.6 m long driver section and a 6.6 m long driven section. It should be noted that the simulation encompasses only the regular length of 6.6 m of the driven section, excluding the additional length equipped with optical diagnostics due to the absence of pressure transducers in that section. By assuming symmetry about the axis of the shock tube, a two-dimensional domain as illustrated in Fig. 2 was used for the simulations. The interior surfaces of the shock tube are modeled as isothermal, maintaining a constant wall temperature of $T_w$ = 294.65 K. The initial pressure in both the driver and driven sections were similar to those used in the experiments, outlined in Table 1. The initial temperature in both the sections was kept as 294.65 K. In the 2D domain, the diaphragm opening process mimics a petal rotating about a hinge on the shock tube wall. At any time instant, the rotation of the petal by an angle $\theta$ results in an orifice of height $h$ as shown in Fig. 2. The petal length, denoted as ($l$), measures 36.2 mm in the HPST. Initially, the value of $h$ is taken as 0.2 mm. For the subsequent time steps, the value of $h$ is determined by the experimentally obtained diaphragm opening profile, as given in Eq.2. The $t_{op}^{exp}$ in Eq. 2 represents the diaphragm opening time and $t_0$ corresponds to time zero. This function controls the opening profile from $t_0$ until the experimentally measured diaphragm opening time. The value of $\theta$ in Eq. 2 is given by Eq. 3, where $A(t)$ indicates the area of the opened aperture at any given time $t$, and $A$ is the total area when fully open.

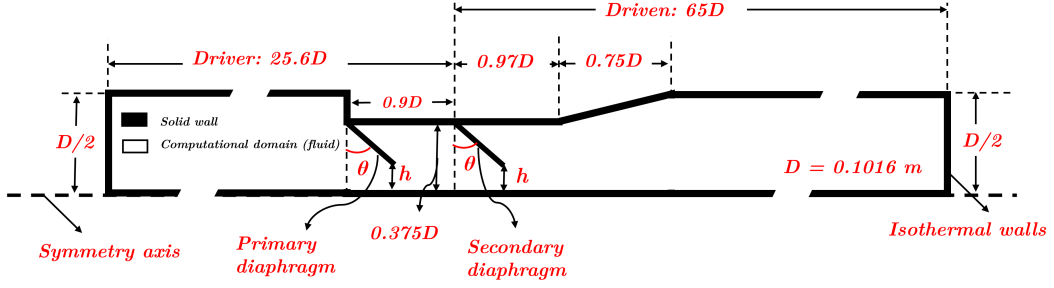

**Fig. 2** Schematic of the numerical domain for the shock tube simulations, illustrating the location of the primary and secondary diaphragms and the dimensions of the various sections of the shock tube. The figure also shows the computational domain (fluid), symmetry axis, and isothermal walls used in the CFD simulations.

$$h = \begin{cases} 0.2\,, mm & \text{if } t = t_0 \\ 0.2 + \ell(1-\cos\theta), mm & \text{if } t_0 < t \leq t_{op}^{exp} \\ 0.2 + \ell\,, mm & \text{if } t > t_{op}^{exp} \end{cases} \quad (2)$$

$$\theta = \cos^{-1}\left(1 - \frac{A_t}{A}\right) \quad (3)$$

Two-dimensional simulations were performed using a commercial CFD software, CONVERGE CFD (v3.1)[34]. A density-based solver was used, employing the PISO scheme for pressure-velocity coupling assuming an ideal gas relation as equation of state. Spatial discretization utilized the $3^{rd}$ Monotonic Upstream-Centered Scheme for Conservation Laws (MUSCL). An adaptive time-stepping strategy governed by diaphragm movement and a Mach number-based Courant-Friedrichs-Lewy (CFL) condition for stability was employed. Converge features a novel cut cell method for mesh generation allowing for simulation of complex and moving boundaries. The base mesh size was set to 2 mm, with an adaptive mesh refinement that reduces cell size to 0.25 mm (minimum cell size $\frac{1}{4}$ base grid/$2^3$). Embedded grids were specifically applied around critical areas such as the diaphragm rupture zone to increase accuracy without substantially increasing computational demands. In a previous study, this grid configuration was found to be adequate for capturing shock wave dynamics with diaphragm opening. [35]. The $k$-$\omega$ SST turbulence model was used, which accurately captures near-wall turbulence dynamics and flow separation, and is robust in shock-driven flow scenarios. Wall shear stress and heat transfer predictions were managed using the Law of the Wall model. The simulations were executed on the SHAHEEN III supercomputer at KAUST, utilizing 1920 cores and completing in approximately 11 hours.

## III. Results

This section elucidates the experimental and numerical results for the operating conditions outlined in Table 1. It is structured into three primary subsections: the first presents shock velocity measurements under varying mid-section pressures and discusses the associated pressure variations. The second subsection discusses the dynamics the diaphragm opening mechanism to elucidate variations in shock velocity in the driven section. The final subsection compares the experimental and numerical results.

**A. Shock Velocity Profiles and Mid-section's Pressure Histories in Experiments**

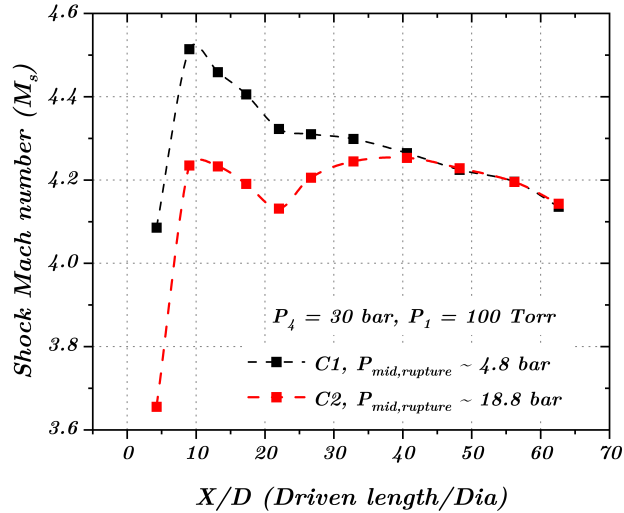

**Fig. 3  Experimentally measured shock velocity profiles in the HPST's driven section for cases C1 and C2 with varying $P_{mid,rupture}$ (4.8 and 1.8 bar) and fixed driver (30 bar) and driven pressures (100 Torr).**

Figure 3 shows the shock velocities measured along the driven section of the shock tube at the conditions specific in Table 1. The normalized driven length, represented on the *x-axis*, is computed by dividing the length by the diameter of the driven section. The computed shock velocities are fit using a modified Bezier curve using Origin Pro software.

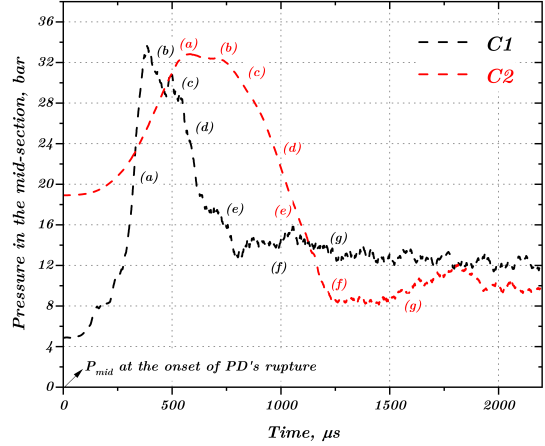

**Fig. 4   Temporal variation in pressure of the mid-section for cases C1 (dashed black line) and C2 (dashed red line), starting from the PD's rupture. Data points (a) through (g) mark the stages in the diaphragm opening process, as illustrated in Figs. 5 and 6.**

At higher driver pressures of 30 bar, as depicted in Fig. 3c, $P_{\text{mid}}$ varied substantially from 4.8 bar to 18.8 bar. As reported by Kashif et al.[28], the shock velocity exhibits a marked dependence on the mid-section pressure. In configuration C1 (with a lower $P_{\text{mid}}$, represented by the black trace) the peak shock Mach number is 4.5 compared to the peak Mach number of 4.2 observed for C2 (with a higher $P_{\text{mid}}$, represented by the red trace). A rapid deceleration phase follows the peak Mach number in both configurations. Furthermore, at around 20 $X/D$, C2 shows a second acceleration phase, lasting until about 40 $X/D$, in contrast to C1, which exhibits a gradual deceleration throughout this region. Remarkably, despite the significant differences in shock velocity histories for C1 and C2, they converge to similar velocities at the end of the driven section.

The key difference in shock velocity profiles, particularly the higher peak Mach number observed in configuration C1 and the secondary acceleration phase in configuration C2, was attributed to two mechanisms in the work of Kashif et al.[28] are now considered. The first mechanism pointed to the delay in rupture of SD due to the interactions of incident shock waves ($ISW_{mid}$) and reflected shock waves ($RSW_{mid}$) in the mid section. This delay results in a higher peak Mach number of the shock, as a high temperature and pressure is produced upstream of the SD due to the shock interactions. The second mechanism is due to the interaction of the reflected shock wave in the mid section ($RSW_{mid}$) with the partially opened PD. This interaction can impede the PD's opening process, potentially slowing down or in some cases, even temporarily reversing the diaphragm's movement. Such dynamics significantly influence the flow rate of the driver gas into the driven section, altering the shock wave velocity. Once the pressure in the mid-section drops due to the SD opening, the driver pressure can facilitate the opening of the PD. This leads to influx of the driver gas into the driven section, contributing to a secondary acceleration phase of the shock wave observed in C2. With the help of pressure measurements, diaphragm imaging and numerical simulations, the present work aims to investigate these two mechanisms in double-diaphragm shock tubes.

The temporal variation in pressure within the mid-section is depicted in Fig. 4 for both configurations C1 (black trace) and C2 (red trace). Time zero marks the onset of the PD's rupture, with initial $P_{\text{mid}}$ values as 4.8 bar for C1 and 18.8 bar for C2. Data points (a) through (g) mark the stages in the diaphragm opening process described in the next section. The rupture of PD generates a shock in the mid-section, initially compressing the gas with an incident shock, followed by a reflected shock on impacting SD. This sequence leads to a pressure rise observable in both cases in Fig. 4. The subsequent pressure drop is a result of reduction in mid-section pressure due to SD rupture. Notably, the pressure jump in C1 is nearly eightfold (from 4.8 to approximately 32 bar) prior to triggering the rupture of SD, while it is just double the initial value in case of C2 (from 18.8 to approximately 32 bar). The larger pressure jump in C1 induces a significant temperature rise as compared to C2. While temperature measurements at these time scales are a challenge in experiments, numerical simulations presented later suggest a temperature difference of approximately 200 K in C1 as compared to C2. This heated slug of gas prior to SD rupture, accelerates the shock wave in the case of C1, thus confirming the first mechanism described by Kashif et al.[28] The second mechanism involving mass flow restriction across the PD can be verified through diaphragm imaging described in the next section.

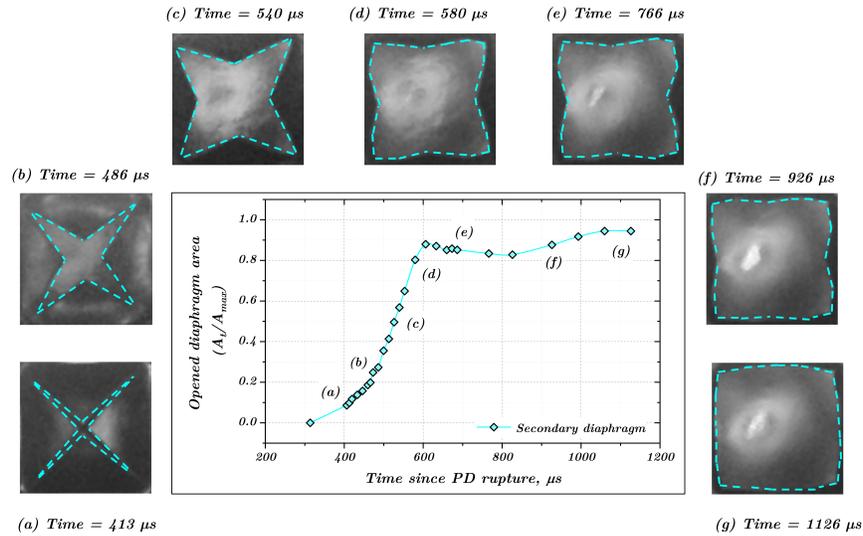

Fig. 5 Sequential visualization of the SD opening dynamics for case C1. Sub-figures (a) through (g) capture the progressive stages of diaphragm rupture and opening at specified timestamps post-primary diaphragm rupture. The plot tracks the normalized area of the diaphragm opening, correlating each image with corresponding data points.

**B. Diaphragm Opening Dynamics**

As mentioned previously, using the imaging setup shown in Fig. 1, the opening dynamics of SD can be captured well. Additionally, the motion of the PD's petals can also be visualized if the SD opens quickly. The images obtained for case C1 are depicted in Fig. 5 labeled (a) through (g) representing different stages of the diaphragm opening. These images sequentially illustrate the progressive opening of the SD over time. Dashed cyan lines show the boundaries of SD and the opened aperture at each stage. The accompanying plot tracks the normalized opened aperture area of SD against the time elapsed since PD rupture. The data points on the plot labeled (a) through (g) correspond to the images with the same labels. At approximately 300 $\mu$s, SD begins to rupture and there is a gradual increase in the opened aperture area. At about 600 $\mu$s, SD has fully opened, allowing the driver gas to flow without any restrictions. After timestamp (e), the opened petals of the diaphragm reflect off the shock tube walls, temporarily reducing the opened aperture area slightly before going back to the fully open state. Figure 6 a similar chronology of events as seen in Fig. 5 is observed for SD although the time instances vary significantly. The rupture of SD starts at approximately 520 $\mu$s until it is fully opened after about 900 $\mu$s. When the SD's opening reaches 40 % of the total final aperture area, the edges of the PD become visible (highlighted with dashed red lines). The red scatter points in the line plot indicate the opened aperture area of the PD. The gradual opening of PD leads to a restricted mass flow of driver gas as compared to configuration C1, resulting in the second acceleration phase of the shock wave in case C2. This experimental evidence supports the hypothesis presented by Kashif et al.[28].

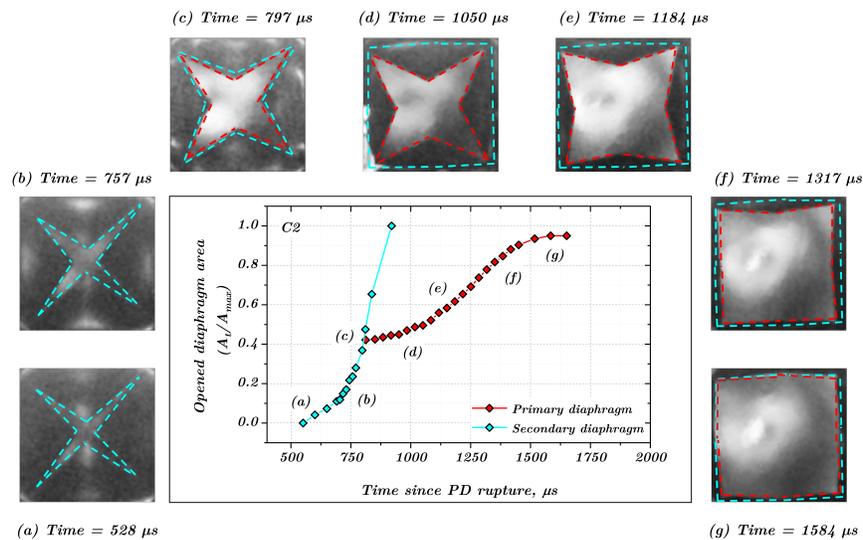

**Fig. 6** **Sequential visualization of the SD opening dynamics for case C2. Sub-figures (a) through (g) capture the progressive stages of diaphragm rupture and opening at specified timestamps post-primary diaphragm rupture. The plot tracks the normalized area of the diaphragm opening, correlating each image with corresponding data points.**

### C. Comparison of Experimental and Numerical Results

In this subsection, the validation of the numerical simulations against the experimentally measured (i) shock velocity profiles and (ii) pressure histories are presented. A brief discussion on the diaphragm opening profiles incorporated in the CFD simulations is provided prior to the validation results.

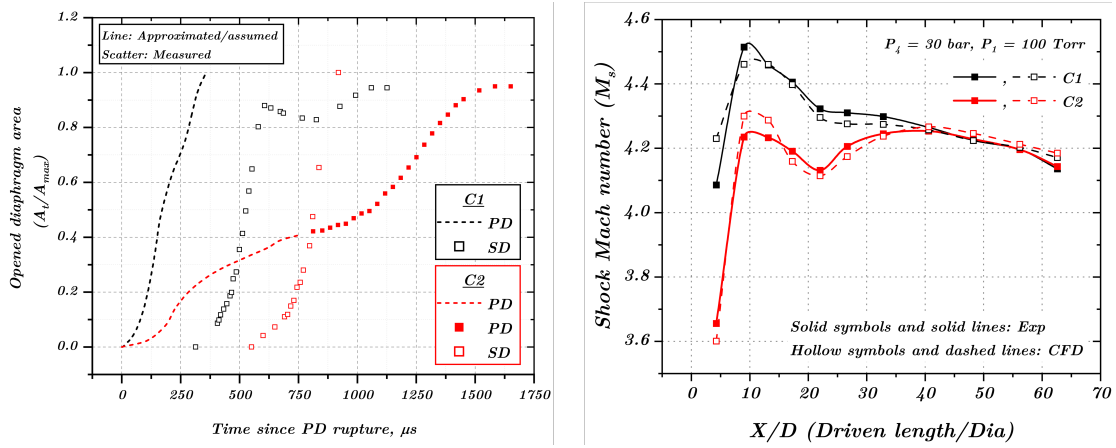

**Fig. 7** (a) Diaphragm opening profiles incorporated in CFD simulations. Data from experiments is shown as scatter points and assumed profiles are shown in dashed lines. Configuration C1 is shown in black while C2 is shown in red. (b) Shock Mach number profiles along the driven section of the shock tube. Solid lines with filled symbols represent experimental data and dashed lines with hollow symbols show CFD results.

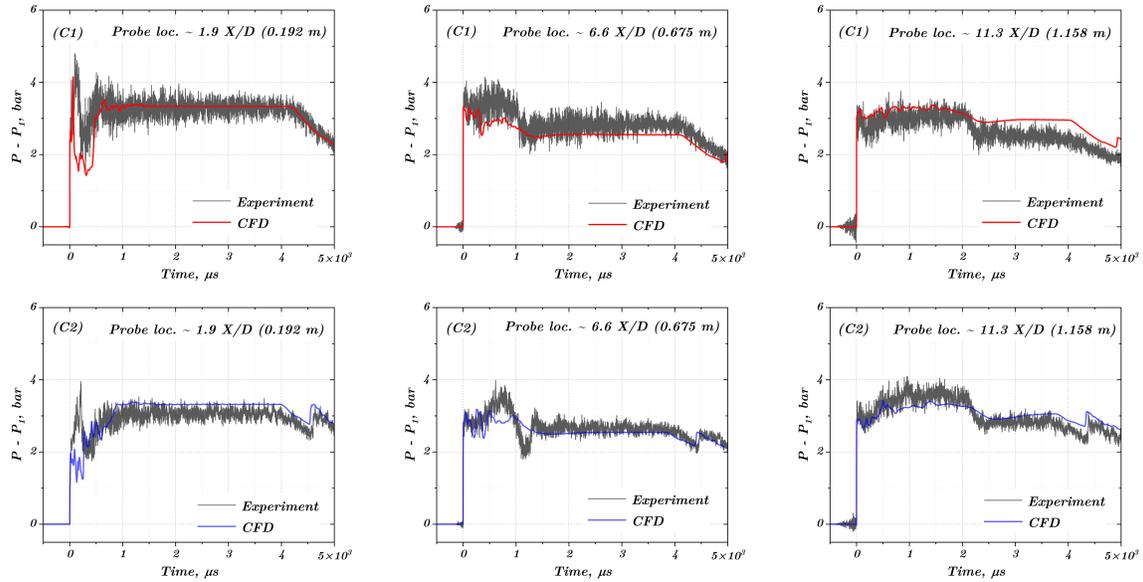

**Fig. 8** Pressure profiles from experiments and CFD simulations at three probe locations for cases C1 and C2. The profiles highlight the comparison at approximately $1.9D$ **(0.192 m)**, $6.6D$ **(0.675 m)**, and $11.3D$ **(1.158 m)**.

As highlighted earlier, experimental measurements of diaphragm opening profiles were available only for SD for both cases. For PD, the opening profiles were absent for case C1, while partial information of opening was available for case C2. Figure 7a depicts the diaphragm opening profiles that were integrated in the CFD simulations; C1 is shown in black and C2 in red. The scatter points represent the experimentally measured diaphragm opening profiles, with dashed lines illustrating the modeled approximations for the PD's openings. For C1, the approximated opening times and profiles for PD were derived from correlations based on single diaphragm experiments (see [35]). Additionally, discussions on the effects of varying PD's opening profiles on shock velocity profiles for case C1 are shown in SM. For C2, with only half of the PD's opening profile available, the remainder profile was approximated. Figure 7b compares shock velocity profiles obtained in experiments and CFD for the two configurations. Solid lines indicate experimental results, while dashed lines represent CFD results. Numerical simulations capture distinct behaviors in C1 and C2, including the higher peak Mach number in C1 and the second acceleration stage in C2, demonstrating the numerical model's capability to replicate complex flow dynamics. Moreover, the simulations accurately predict the similar shock Mach numbers at the end of the driven section for both configurations.

Figure 8 shows the comparison of pressure histories from CFD simulations and experimental measurements at the first three transducer locations for both cases C1 and C2. These plots illustrate the temporal pressure profiles measured by probe locations at $1.9D$ (0.192 m), $6.6D$ (0.675 m), and $11.3D$ (1.158 m). The black lines represent experimental data, while the red lines depict numerical results for C1 and blue lines for C2. In case C1, the simulation closely aligns with experimental observations at all probe locations, demonstrating accurate replication of the pressure dynamics. Case C2, on the other hand, shows slight discrepancies for the pressure sensor at $1.9D$, particularly evident in the initial shock response. This could be attributed to the approximation in the approximation in PD's opening profile. Overall, the pressure jumps and arrival of expansion waves at all the pressure sensors are captured well for both the cases. This comparison validates the CFD model's effectiveness in mimicking the shock velocity formation and propagation in a double diaphragm shock tube.

## IV. Discussion

This section elucidates the findings from numerical simulations conducted with dual objectives: firstly, to accurately replicate the shock profiles observed in cases C1 and C2, specifically addressing the delayed rupture of the SD and the intermittent opening processes of the PD; and secondly, to quantify the axial gradients in various thermodynamic properties resulting from variations in shock velocity. These simulations are crucial for validating the proposed mechanisms that govern shock velocity variations and for examining their implications in double-diaphragm shock tube configurations. The discussion is structured into two parts: the initial subsection explores the flow evolution instigated by the ruptures of the diaphragms; and the second subsection delves into the quantification of axial gradients in temperature and pressure within the heated slug for both cases C1 and C2.

### A. Flow Evolution Near the Diaphragm Station

This subsection delves into the complex interplay of flow dynamics and temperature variations observed near the diaphragm and extending approximately up to 0.8 meters ($8D$) into the driven section of the shock tube. Detailed insights are provided through temperature contours and numerical schlieren visualizations as depicted in Figs. 9 and 10, respectively, for configurations C1 and C2. These visual representations clarify the observations of flow phenomena following diaphragm rupture, as well as offer a comparative analysis of the distinct flow characteristics inherent to each configuration, thereby accentuating both similarities and differences in shock behavior. The timestamps next to the contours indicate the time instant post-rupture of PD.

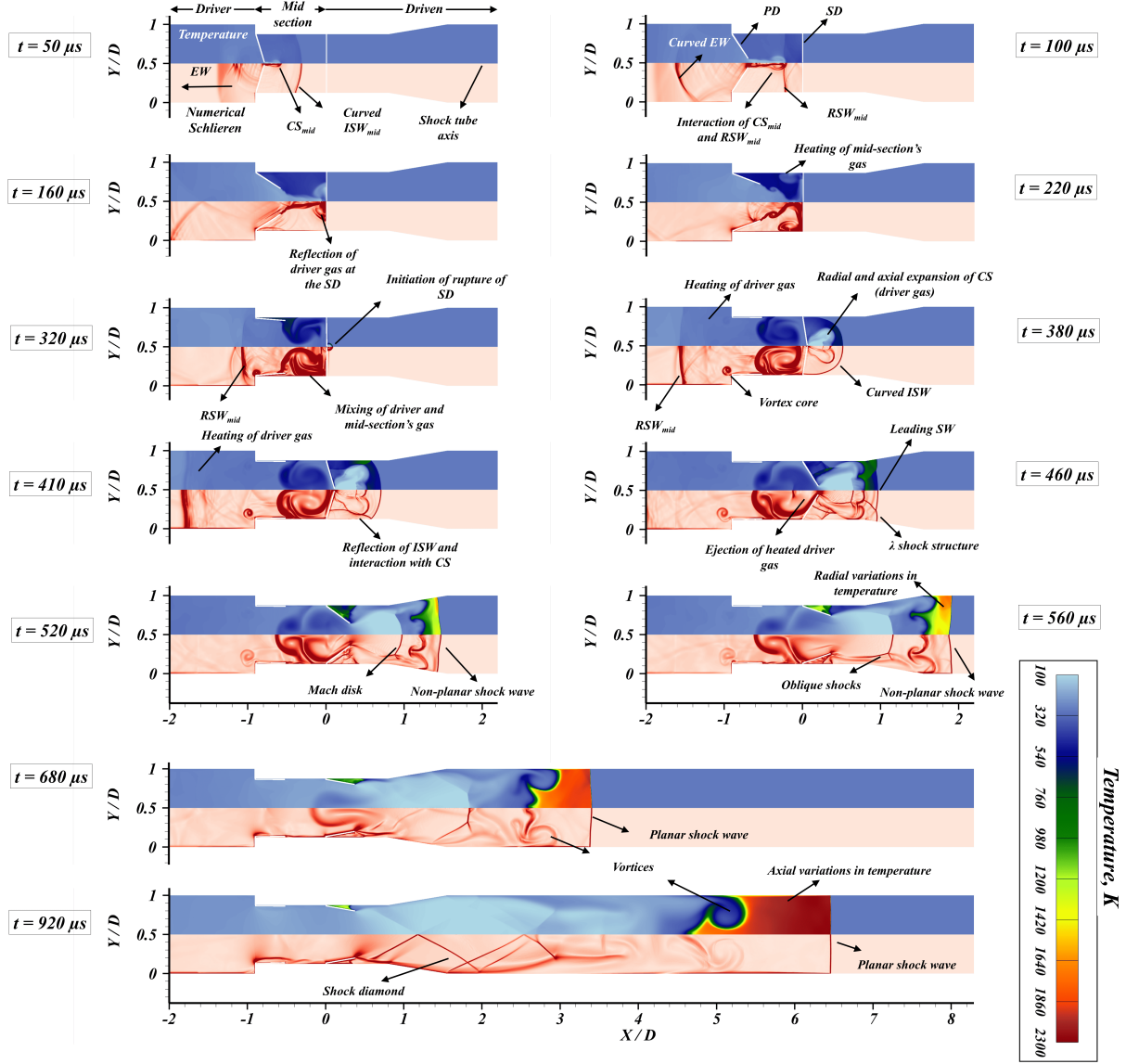

Fig. 9 Temporal evolution of flow features and temperature gradients in Case C1. The series of contours illustrate key phenomena such as expansion wave interactions, diaphragm ruptures, and complex shock wave dynamics at various times post-rupture of the diaphragms, highlighting the evolution of shock behavior and thermal effects within the shock tube.

At $t = 50$ μs, a curved intermediate shock wave ($ISW_{mid}$) is formed in the mid-section, as evident in the initial contours of Figs. 9 and 10 in both configurations. Moreover, the shock wave is of a lower strength for case C2. There is a small temperature rise behind the shock front in both C1 and C2. Additionally, the expansion of the driver gas in the mid-section, denoted by $CS_{mid}$, is distinctly visible. At $t = 100$ μs in C1 and $t = 120$ μs in C2, the $ISW_{mid}$ undergoes a reflection at the SD, resulting in the formation of a reflected shock wave ($RSW_{mid}$). This leads to a further temperature and pressure rise in the mid-section. Since $RSW_{mid}$ is weaker in C2 as compared to C1, the pressure and temperature rise in C1 significantly higher than that observed in C2. Localized regions of elevated temperature emerge in the mid-section in the case of C1. An almost 200 K increase is seen at $t = 220$ $\mu$s in Fig. 9. Conversely, the temperature rise relatively lower in the case of C2.

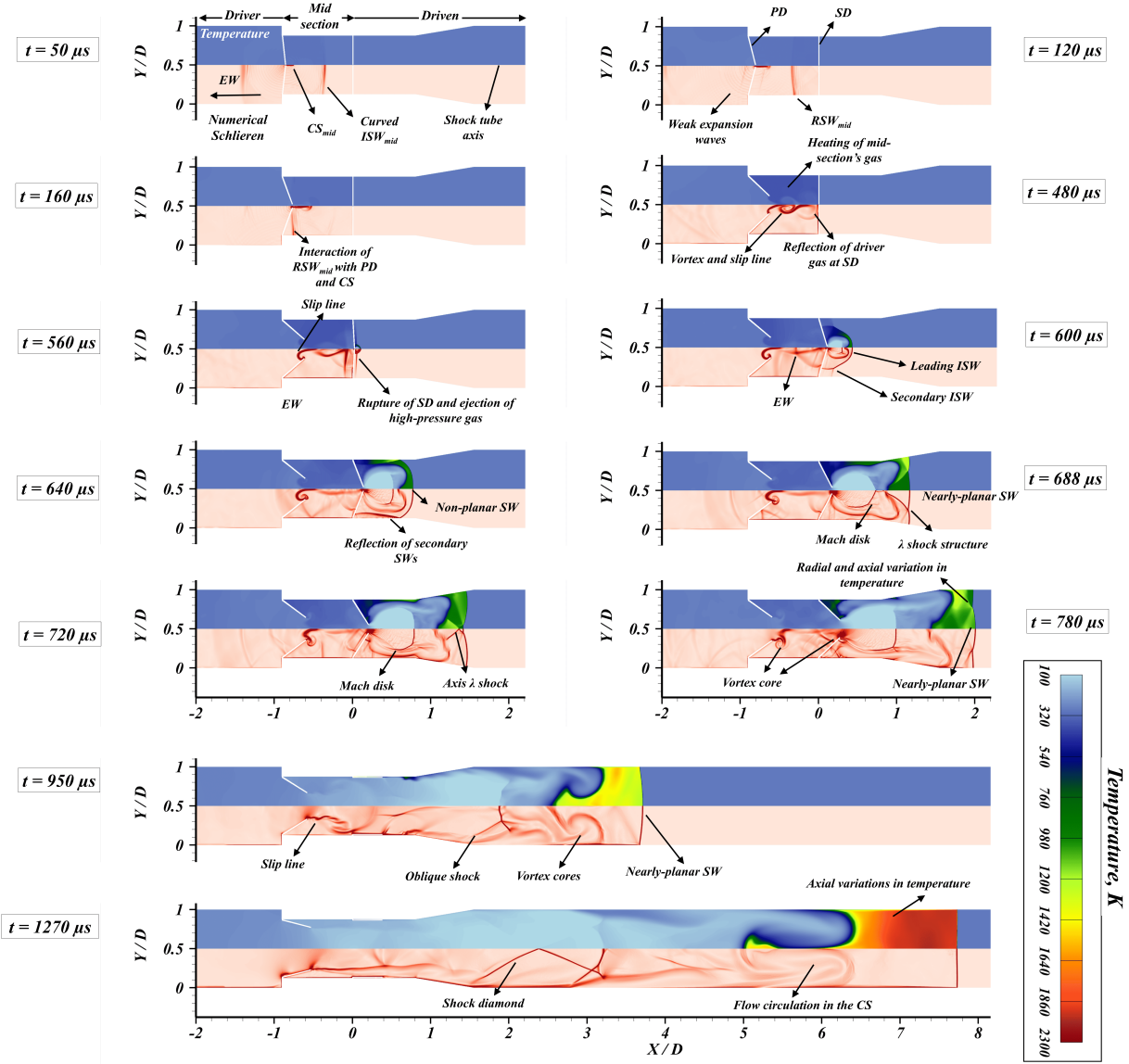

**Fig. 10** Temporal evolution of flow features and temperature gradients in Case C2 similar to Fig. 9

The third row of contours showcase the rapid expansion of the gas in the mid-section upon SD's rupture, generating a curved shock wave ($ISW_{driven}$). The shock heated gas behind the SD expands into the driven section. In the case of C1, the shock heated gas expands quicker due to the higher temperature, hence accelerating the shock wave to a higher speed as compared to case C2. An important observation in the case of C1 is the propagation of $RSW_{mid}$ upstream into the driven section, which induces a mild temperature increase in the driver gas — a phenomenon nearly absent in the case of C2. The fourth and fifth rows of contours in Figs. 9 and 10 showcase various flow features such as the expansion of the mid-section gas and the interaction of the incident shock wave with the shock tube wall. Multiple $\lambda$ shock structures are visible near the tube wall and at the center-line of the shock tube. Additionally, the expanding high-pressure gas forms a Mach disc within the mid-section gas, with multiple vortex structures seen in both configurations. C1 and C2 exhibit different levels of radial and axial temperature variations in the heated driven gas behind the incident shock wave, due to the complex flow expansion of the mid-section gas.

In the final two contours, the temperature and density begin to equalize in the radial direction ,dominated by thermal diffusion, leading to a more uniform distribution of temperature and density contours as observed at $t = 920$ $\mu$s for C1 and $t = 1270$ $\mu$s for C2. However, the axial temperature equalization is significantly slower due to longer length scales requiring several hundred microseconds to take effect. This slow process results in marked axial temperature variations for both configurations, with significant differences between C1 and C2. The following section provides quantitative comparisons of temperature and pressure distributions along the center-line in the shocked gas, enhancing the understanding of these physical phenomena.

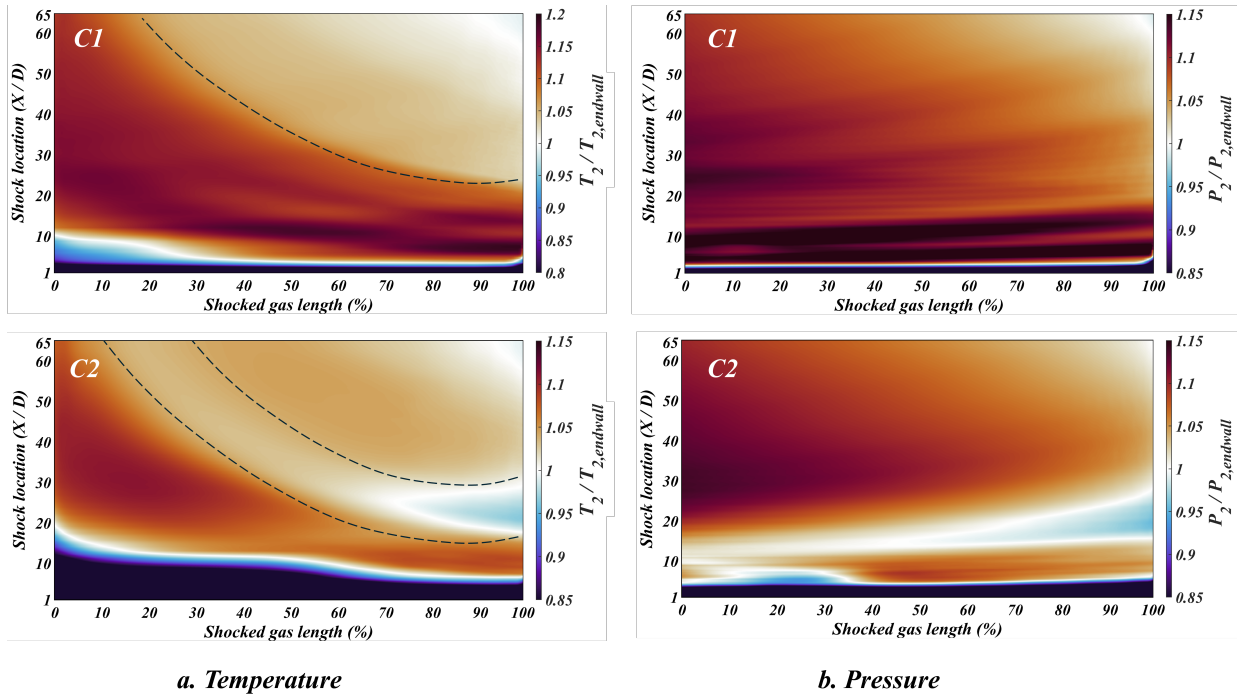

*a. Temperature*        *b. Pressure*

**Fig. 11 Axial temperature and pressure variation in the heated shocked gas as a function of shock location *(y-axis)* for both C1 and C2. The *x-axis* in these plots represent the normalized length of the driven gas behind the shock wave, 100% represents the gas slug immediately behind the shock wave, and 0% represents the gas slug immediately ahead of the contact surface. The *y-axis* represents the shock location in the driven section normalized by diameter. Dashed lines are added to support the discussion.**

**B. Axial Gradients of Temperature and Pressure in the *Shocked* Gas**

Axial variations in thermodynamic parameters in the shocked gas are a result of shock velocity variations in the driven length of the shock tube. The differences in the shock profiles between C1 and C2 (Fig. 7b) is expected to influence the thermodynamic state of the *shocked* driven gas. To quantify these variations, center-line temperature and pressure contours from CFD are presented in Fig. 11. The *x-axis* in these plots represent the normalized length of the driven gas behind the shock wave, 100% shocked gas length represents the gas slug immediately behind the shock wave, and 0% represents the gas slug immediately ahead of the contact surface. The *y-axis* represents the shock location in the driven section normalized by diameter, with 1 $X/D$ being close to the diaphragm, and 65 $X/D$ being close to the driven end wall. The temperature and pressure values are normalized against the reference values obtained from the shock velocity at the end wall and ideal shock relations to emphasize the differences more clearly.

**Temperature variation:** The temperature of the shock heated gas is significantly influenced by the shock velocity variation. When the shock wave is initially close to the diaphragm location, it has a low velocity compared to its peak value and therefore, the initial slug of gas is at lower temperatures (< 10 X/D for C1 and < 20 X/D for C2 in Fig. 11a). The shock wave begins to accelerate downstream of the diaphragm location and the gas slug, initially at a lower temperature, eventually gets replaced by hotter shocked gas. For C1, the peak velocity occurs around 10 X/D (as seen in Fig. 7b), at which point the shocked gas reaches its highest temperature. Subsequently, the shock wave begins to decelerate due to boundary layer growth and therefore, cooler gas fills the shocked gas region. in C1, this effect is clearly seen from 30 X/D in Fig. 11a, where the black dashed line shows the continuous increase in the concentration of a cooler gas being added. When the shock wave reaches the end wall, about 75 % of the shocked gas has a relatively lower temperature than the remaining 25 %. On the contrary, the temperature profile of the shocked gas for C2 is influenced by the second stage of shock acceleration occurring in the region between 15 and 35 $X/D$. In this region, the introduction of cooler gas is evident due to the initial shock deceleration from 12 $X/D$ (refer to Fig. 7b). This cooler gas gradually diminishes as it heats up, influenced by diffusion and the progressive acceleration of the shock from 22 $X/D$ onward. The dashed lines in Fig. 11a (for C2) indicate the continuous progression of this perturbed temperature state until the shock reaches the endwall. Boundary layer effects similar to those observed in C1 also become evident.

Given the similarity in shock velocities and resultant temperatures at the endwall for both C1 and C2, directly comparing the temperature profiles across configurations is informative. As shown in Fig. 12a, at an earlier shock location of 13.2 $X/D$, the temperature profiles differ significantly between C1 and C2. C1 exhibits a peak temperature rise of over 17%, compared to less than 7% in C2. As the shock progresses downstream, these differences in temperature profiles diminish notably. At the instant the shock reaches the end wall, the gas mixtures for both configurations exhibit similar temperature distributions across the last 60% of the shocked gas length from the end wall, attributable to comparable final attenuation rates. However, within the initial 40% of the shocked gas region, the temperature differences remain pronounced, with peak temperatures for C1 being 12.5% higher and for C2 being 7% higher, due to the initial steep variations in shock velocity between the two configurations.

**Pressure variation:** Pressure variations within the shocked gas exhibit a distinct characteristic pattern compared to temperature due to the inherently fast relaxation of pressure disturbances. In both configurations, C1 and C2, pressure changes rapidly adjust to new equilibrium states following shock interactions. This process is significantly faster than temperature equalization due to the high propagation speed of pressure waves (sound waves). The contours depicting normalized pressure variations along the shocked gas length as a function of shock location are shown in Figure 11b. Notably, in C2, the pressure dynamics are influenced by the second stage of acceleration. However, these variations stabilize rapidly, and minimal differences in pressure are observed between C1 and C2, as demonstrated in Figure 12b. Quantification of the pressure variation in Figure 12b reveals a consistent linear pressure trend along the shocked gas length, with up to 10% higher pressure relative to the end wall when the shock reaches the driven end wall.

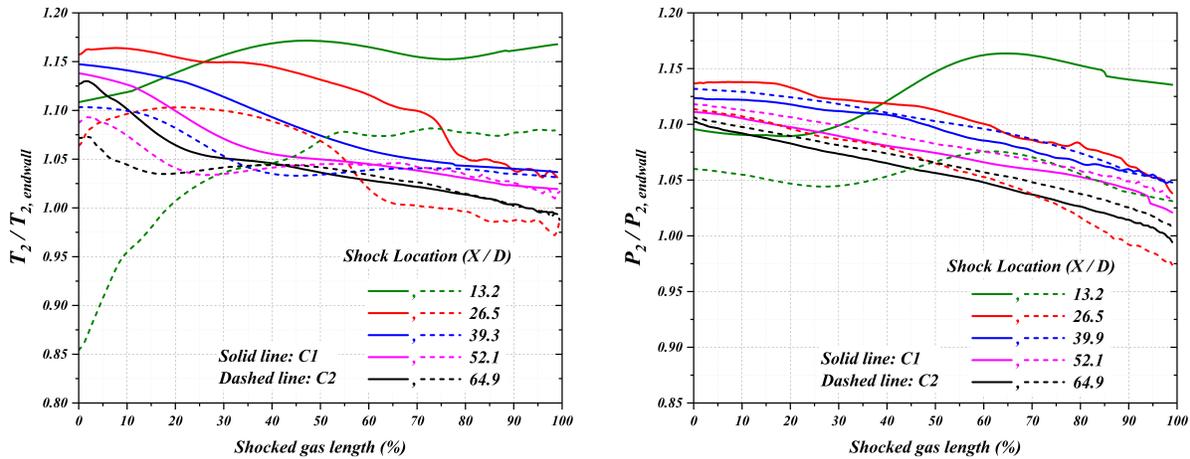

Fig. 12  Comparison of axial variation of temperature and pressure profiles in the heated shocked gas for C1 and C2. The *x-axis* in these plots represent the normalized length of the driven gas behind the shock wave, 100 represents the gas slug immediately behind the shock wave, and 0 represents the gas slug immediately ahead of the contact surface

## V. Conclusions

This study investigates the dynamics of shock wave propagation in double-diaphragm shock tubes through combined experimental measurements and numerical simulations, offering new insights into the complex flow phenomena within such configurations. By examining cases with varying mid-section pressures, the study explores the impact of diaphragm dynamics and mid-section pressure variations on shock velocity profiles, axial thermodynamic gradients, and the evolution of flow features. Key findings highlight the significant role of mid-section pressure in influencing shock velocity and summarized below:

- Experimental results demonstrate that lower mid-section pressures facilitate higher peak shock velocities and a more rapid deceleration phase, while higher mid-section pressures induce a second stage of acceleration, corroborating previous observations.

- Visualization of diaphragm-opening dynamics reveals critical interactions between the reflected shock wave in the mid section region and the primary diaphragm, validating the mechanisms governing shock velocity variations.

- Numerical simulations replicate experimental shock profiles with high accuracy and provide detailed insights into the flow evolution near the diaphragm station. They show that heated gas expelled from the mid-section significantly influences axial temperature and pressure gradients within the shocked gas.

- Case C1 (lower *Pmid*) exhibits steeper axial temperature gradients due to higher initial shock velocities, while case C2 (higher *Pmid*) shows a more gradual temperature profile influenced by secondary acceleration. In contrast, pressure equalization occurs much faster, with minimal differences observed between the two cases, highlighting the distinct relaxation dynamics of pressure and temperature.

These findings underscore the necessity of precise control over mid-section pressure and diaphragm dynamics to minimize axial thermodynamic variations in shocked gas. Such control is crucial for enhancing the uniformity of flow conditions in shock tube experiments, particularly for applications in chemical kinetics and combustion studies. Insights into flow features—such as shock wave interactions, Mach disk formations, and boundary layer effects—provide a foundation for advancing the design and operation of double-diaphragm shock tubes.

Future work will extend these analyses to reflected shock regions, focusing on boundary layer growth and its interaction with the reflected shock wave. Additionally, integrating experimental diagnostics, such as laser absorption spectroscopy, with CFD simulations to quantify temperature variations could further enhance understanding of shock tube dynamics and their implications for diverse research applications.

## Acknowledgments

This work was sponsored by King Abdullah University of Science and Technology (KAUST) and supported by the KAUST Supercomputing Laboratory (KSL). All simulations were performed on KSL's Shaheen III supercomputer. Convergent Science provided CONVERGE licenses and technical support for this work.## References